\documentclass[twocolumn,amsmath,floatfix]{revtex4}
\usepackage{latexsym}
\usepackage{amssymb}
\usepackage{pdfpages}
\usepackage{graphics}
\usepackage{ulem}

\begin{document}
\newcommand{\ja}{Jakuba\ss a-Amundsen }
\newcommand{\bfx}{\mbox{\boldmath $x$}}
\newcommand{\bfq}{\mbox{\boldmath $q$}}
\newcommand{\bfnabla}{\mbox{\boldmath $\nabla$}}
\newcommand{\bfalpha}{\mbox{\boldmath $\alpha$}}
\newcommand{\bfsigma}{\mbox{\boldmath $\sigma$}}
\newcommand{\bfeps}{\mbox{\boldmath $\epsilon$}}
\newcommand{\bfA}{\mbox{\boldmath $A$}}
\newcommand{\bfP}{\mbox{\boldmath $P$}}
\newcommand{\bfe}{\mbox{\boldmath $e$}}
\newcommand{\bfn}{\mbox{\boldmath $n$}}
\newcommand{\bfW}{{\mbox{\boldmath $W$}_{\!\!rad}}}
\newcommand{\bfM}{\mbox{\boldmath $M$}}
\newcommand{\bfI}{\mbox{\boldmath $I$}}
\newcommand{\bfJ}{\mbox{\boldmath $J$}}
\newcommand{\bfQ}{\mbox{\boldmath $Q$}}
\newcommand{\bfY}{\mbox{\boldmath $Y$}}
\newcommand{\bfp}{\mbox{\boldmath $p$}}
\newcommand{\bfk}{\mbox{\boldmath $k$}}
\newcommand{\bfks}{\mbox{{\scriptsize \boldmath $k$}}}
\newcommand{\bfqs}{\mbox{{\scriptsize \boldmath $q$}}}
\newcommand{\bfxs}{\mbox{{\scriptsize \boldmath $x$}}}
\newcommand{\bfalphas}{\mbox{{\scriptsize \boldmath $\alpha$}}}
\newcommand{\bfs}{\mbox{\boldmath $s$}_0}
\newcommand{\bfv}{\mbox{\boldmath $v$}}
\newcommand{\bfw}{\mbox{\boldmath $w$}}
\newcommand{\bfb}{\mbox{\boldmath $b$}}
\newcommand{\bfxi}{\mbox{\boldmath $\xi$}}
\newcommand{\bfzeta}{\mbox{\boldmath $\zeta$}}
\newcommand{\bfr}{\mbox{\boldmath $r$}}
\newcommand{\bfrs}{\mbox{{\scriptsize \boldmath $r$}}}
\newcommand{\bfps}{\mbox{{\scriptsize \boldmath $p$}}}

\renewcommand{\theequation}{\arabic{section}.\arabic{equation}}
\renewcommand{\thesection}{\arabic{section}}
\renewcommand{\thesubsection}{\arabic{section}.\arabic{subsection}}

%command to highlight Xavi's comments in blue
\newcommand{\xrm}[1]{{\color{blue} #1}}
\newcommand{\can}[1]{{\color{red} \sout{#1}}}
\newcommand{\com}[1]{{\color{brown} #1}}

\title{\Large\bf Dispersion effects in elastic electron scattering from  $^{208}$Pb}

\author{D.~H.~Jakubassa-Amundsen$^1$ and X.~Roca-Maza$^2$\\
$^1$Mathematics Institute, University of Munich, Theresienstrasse 39,\\ 80333 Munich, Germany\\
$^2$Dipartimento di Fisica, Universit\'{a} degli Studi di Milano and INFN, Sezione di Milano, \\
Via Celoria 16, 20133 Milano, Italy}

%\date{}
%\date{\today}
%\maketitle

\vspace{1cm}

\begin{abstract}  
Dispersion corrections to elastic electron scattering
from $^{208}$Pb at energies up to 150 MeV are estimated within the second-order Born approximation.
All strong transient nuclear excited states 
with angular momentum up to $L=3$ and natural parity are taken into consideration.
It is found that at small scattering angles the high-lying $L=1$ isovector states provide the dominant  contribution to the dispersive change of the differential cross section
and of the beam-normal spin asymmetry. 
At the backmost angles these changes are considerably smaller and are basically due to the isoscalar $L=2$ and, to a lesser extent, $L=3$ states.
In comparison with the findings for $^{12}$C,  the dispersive cross-section modifications for $^{208}$Pb are significantly larger, while the spin-asymmetry changes are much smaller at all angles. The latter originates from the destructive interference of the multipole contributions in $^{208}$Pb that switch sign when proceeding from $L$ to $L+1$. This mechanism might be  the origin of the small beam-normal spin asymmetries recently measured in ${}^{208}$Pb at much larger energies.

\end{abstract}

\maketitle

\section{Introduction}

The influence of transiently excited nuclear states in high-energy elastic electron-nucleus scattering has attracted a great deal of interest lately, both experimentally \cite{We01,Ab12,Es18,Jeff20,Ad21} and theoretically \cite{GH08,AM04}.
A yet unresolved problem concerns the influence of these dispersion effects on the spin asymmetry for electrons polarized perpendicular to the beam axis.
Such experiments were carried out at GeV collision energies and small scattering angles, and
in the case of $^{12}$C and other light target nuclei
 they could be explained by calculations profiting from the relation between dispersion and the experimental forward Compton scattering cross section 
 \cite{Ab12, Es18}.
However, the measured asymmetry for $^{208}$Pb was more than an order of magnitude smaller than predicted by this theory \cite{Ab12,Ad21}.
This discrepancy could not even be remedied by applying, instead of the second-order Born approximation, a nonperturbative treatment of the dispersion effects \cite{Ko21}.

Another topic of interest concerns precision experiments at beam energies between $50-150$ MeV and larger scattering angles, to be performed at the Mainz MAMI and MESA facilities in the near future.
Carbon targets are envisaged for test studies concerning parity violation \cite{Au18,Go22}, where the occurrence of a beam-normal spin asymmetry is an unwanted background effect. 
Also the use of lead or gold targets is planned for low-energy Mott polarimetry in order to determine the degree of beam polarization for the MESA experiments \cite{Tha22}.

In the present work we investigate whether the nuclear dipole excitations which dominate the photoabsorption cross section \cite{Ve70,Da92} --as employed in \cite{GH08,Ko21} at large collision energies-- really govern the dispersion effects for the lead nucleus as they do for $^{12}$C at electron beam energies between tenths and hundreds of MeV and for a large range of scattering angles.

In order to estimate the influence of dispersion on elastic electron scattering, 
 early work \cite{Sch55,Le56},  particularly for $^{12}$C \cite{FR74}, introduced closure approximations for coping with the sum over the nuclear excitations.
An explicit inclusion of transiently excited nuclear states was carried out  by De Forest \cite{Fo70} and Bethe and Molinari \cite{BM71} within the harmonic oscillator model.
More advanced calculations considered a few excited nuclear states within a close-coupling formalism \cite{MR74}, or within the related eigenchannel theory \cite{TG69} using collective nuclear models.

In the prescription given below, both isoscalar and isovector  excitations of the $^{208}$Pb nucleus up to an angular momentum $L=3$ are taken into consideration 
within the Born approximation. The required charge and current transition 
densities for these excited states are calculated within a self-consistent Hartree-Fock (HF) plus Random Phase Approximation (RPA) \cite{Co13,Co21}, successfully used in the literature to describe the nuclear structure and collective excitations \cite{Be03,Pa07,Na16}.
 The method is based on the Skyrme parametrization SkP \cite{Do84} which provides the correct level density of states at the Fermi surface. This choice helps to furnish
 a reasonable description of collective excited states up to the Giant Resonance region, that is, those expected to dominate in our analysis.
By investigating a total number of ten intermediate states, up to an excitation energy of 30 MeV, the assumption \cite{BM71,BC72} is tested also here  whether the dipole states in the giant resonance region contribute predominantly to the dispersive changes of the spin asymmetry.
For the carbon nucleus, this conjecture could be verified in the case of not too high momentum transfer, by considering three prominent states with $L\leq 2$ \cite{Jaku23}.

The paper is organized as follows.
Section 2 contains a brief description of the theory and a presentation of the form factors for the dominant nuclear excitations. In section 3 results for the dispersive cross-section and spin-asymmetry changes at impact energies of 56 MeV and 150 MeV are provided.
A comparison with the results for the $^{12}$C target is given as well.
The conclusion is drawn in section 4. Atomic units ($\hbar=m=e=1)$ are used unless indicated otherwise.

\section{Theory}

Only dispersion effects will be considered here,
while the other  radiative corrections are discussed e.g. in \cite{T60,MT00}.
When, to lowest order, the two-photon-exchange or box-diagram contribution is included,
 the differential cross section for electrons, spin-polarized in the direction $\bfzeta_i$ and
scattering elastically into the solid angle $d\Omega_f$, is given by
\begin{equation}\label{2.1}
 \frac{d\sigma_{\rm box}}{d\Omega_f}(\bfzeta_i)=\, \frac{|\bfk_f|}{|\bfk_i|}\;\frac{1}{f_{\rm rec}} \sum_{\sigma_f} \left[\, |f_{\rm coul}|^2 
 +\,2\mbox{ Re }\{f_{\rm coul}^\ast A_{fi}^{\rm box}\} \right],
\end{equation}
where $\bfk_i$ and $\bfk_f$ are, respectively, the momenta of ingoing and scattered electron.
The  transition amplitude $f_{\rm coul}$ describes the leading-order potential scattering by the nuclear target field $V_T$,
from which the  (Coulombic) cross section is derived,
\begin{equation}\label{2.2}
\frac{d\sigma_{\rm coul}}{d\Omega_f}(\bfzeta_i)\,=\,\frac{|\bfk_f|}{|\bfk_i|}\,\frac{1}{f_{\rm rec}}\;\sum_{\sigma_f}|f_{\rm coul}|^2.
\end{equation}
A sum over the spin projections $\sigma_f$ of the scattered electron is included.
Recoil is accounted for in $f_{\rm coul}$ by a reduced collision energy $\bar{E}=\sqrt{(E_i-c^2)(E_f-c^2)}$ and by the kinematical recoil factor
\begin{equation}\label{2.3}
f_{\rm rec}\,=\,1\,-\,\frac{q^2\,E_f}{2M_Tc^2\,\bfk_f^2}\;\left(1-\,\frac{c^2}{M_TE_f}\right),
\end{equation}
where $q=((E_i-E_f)/c,\bfk_i-\bfk_f)$ is the 4-momentum transfer to the target nucleus, with $E_i$ and $E_f$ the total energies of initial and final electron, respectively,
and $M_T$ is the target mass number.
The dispersion amplitude $A_{fi}^{\rm box}$ is calculated by means of  the Feynman box diagram in second-order Born theory \cite{BD64,FR74},
$$A_{fi}^{\rm box}\,=\,\frac{\sqrt{E_iE_f}}{\pi^2c^3}
\sum_{LM;\omega_L\neq 0} \int d\bfp$$
\begin{equation}\label{2.4}
\times\;\sum_{\mu,\nu=0}^3 \frac{1}{(q_2^2+i\epsilon)(q_1^2+i\epsilon)}\;t_{\mu\nu}(p)\;T^{\mu\nu}(LM,\omega_L).
\end{equation}
The denominator results from the two-photon propagator with $q_1=k_i-p$ and $q_2=p-k_f$ where $p=(E_p/c,\bfp)$ is the 4-momentum of the electron in its intermediate state,
which enters into the electronic transition matrix element $t_{\mu\nu}$,
$$t_{\mu \nu}(p)\;=\;c\,u_{k_f}^{(\sigma_f)+}\gamma_0 \gamma_\mu\,\frac{E_p+c\bfalpha \bfp + \beta mc^2}{E_p^2-\bfp^2c^2-m^2c^4+i\epsilon}$$
\begin{equation}\label{2.5}
\times \;\gamma_0\gamma_\nu\;u_{k_i}^{(\sigma_i)}.
\end{equation}
In accordance with the second-order Born approximation, the electron is represented by the free 4-spinors $u_{k_i}^{(\sigma_i)}$ and $u_{k_f}^{(\sigma_f)}$
relating to the spin projections $\sigma_i$ and $\sigma_f$, respectively. The symbols $\bfalpha, \;\beta,\;\gamma_\mu, \,\mu=0,...,3$ are Dirac matrices \cite{BD64}.

\begin{figure}
%\vspace{0.5cm}
\includegraphics[width=7cm]{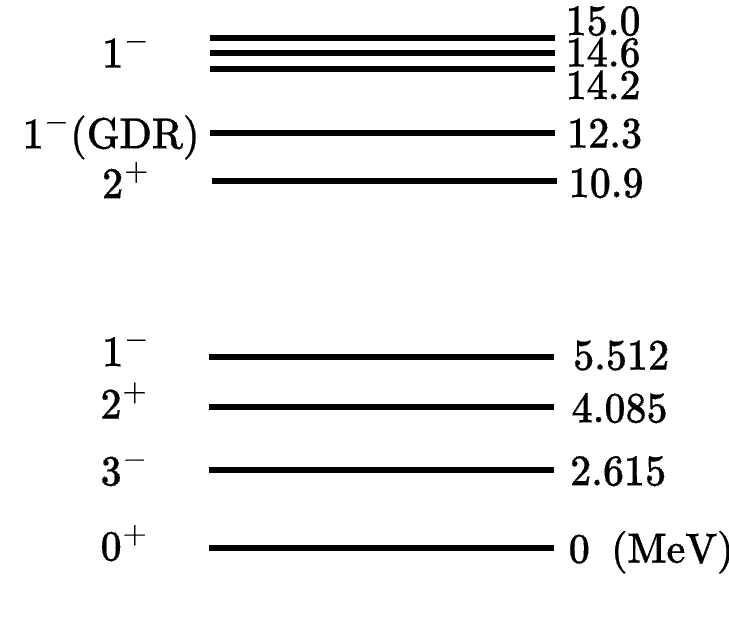}
%\vspace{-1.5cm}
\caption
{
Truncated level scheme of the $^{208}$Pb nucleus (see text for details).
Shown are the strong $L^\pi$ states with $L\leq 3$
(together with their excitation energies up to 15 MeV) that are included in the calculation of dispersion.
}
\label{fig1}
\end{figure}

The nuclear transition matrix element $T_{\mu\nu}$ accounts for the virtual excitation of the nucleus into the state $|LM,\omega_L\rangle$ with angular momentum $L$,
magnetic projection $M$ and energy $\omega_L$.
For $^{208}$Pb, the predominantly excited states of low angular momentum ($L\leq 3$) are shown in Fig.~\ref{fig1}.  For illustrative purposes, the energies of the lower four states are adopted from the measurements, while the energies of all higher-lying states
are taken from theory \cite{Pa07}: the present SkP HF+RPA and the quasiparticle phonon  model \cite{JP16}. It is to be noted that we are only considering nuclear states with natural parity since in general such electric transitions are much
stronger than the magnetic excitations of the odd-parity states for the kinematics considered here.

%\vspace{1.2cm}

%Fig.1\\

%\vspace{0.5cm}

Thus $T_{\mu\nu}$ describes the excitation of the ground-state  nucleus $|0\rangle$ and its subsequent decay into the ground state, mediated  by  the transition density operator $\hat{J}=(\hat{\varrho},\hat{\bfJ})$,
\begin{equation}\label{2.6}
T_{\mu\nu}(LM,\omega_L)\,=\,\langle 0|\,\hat{J}_\mu(\bfq_2)|LM,\omega_L\rangle \langle LM,\omega_L|\,\hat{J}_\nu(\bfq_1)|0\rangle.
\end{equation}

%Fig.2
\begin{figure*}[t]
\centering
\begin{tabular}{cc}
\hspace{-1cm}\includegraphics[width=.7\textwidth]{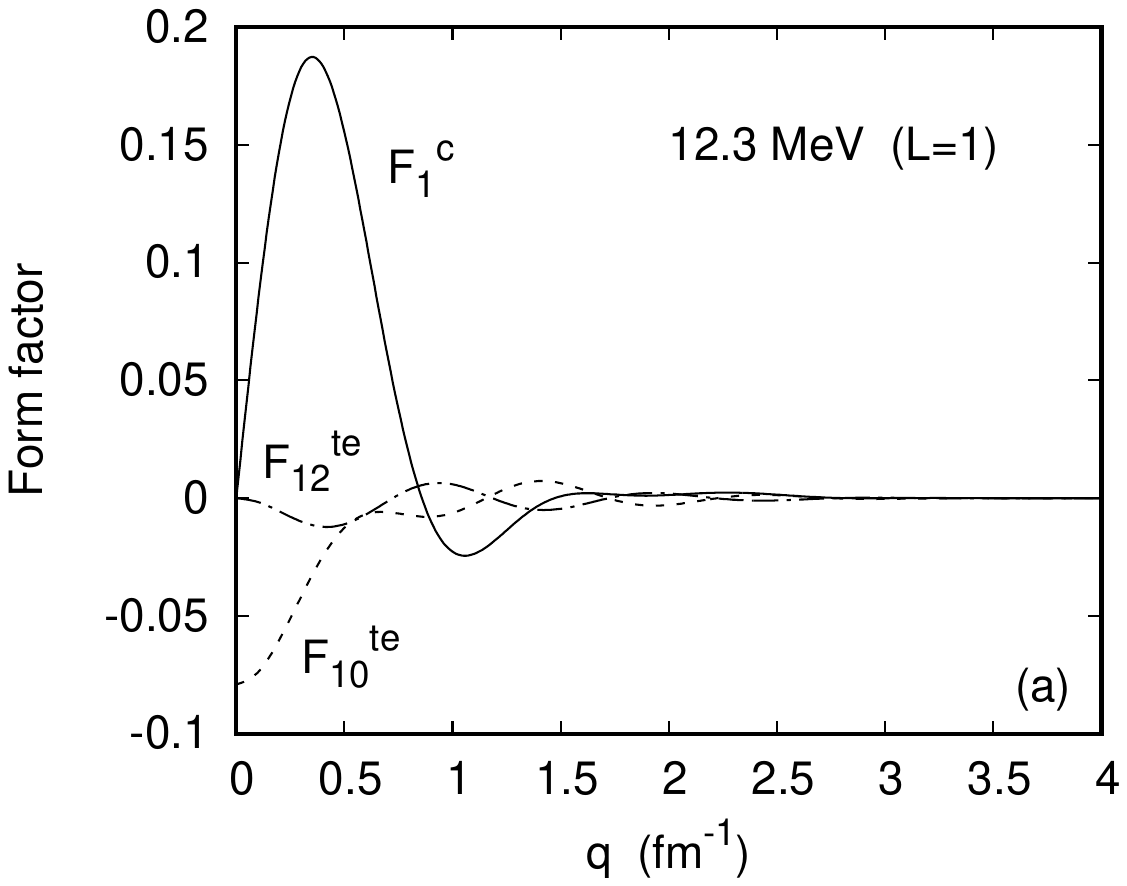}&
\hspace{-3.0cm} \includegraphics[width=.7\textwidth]{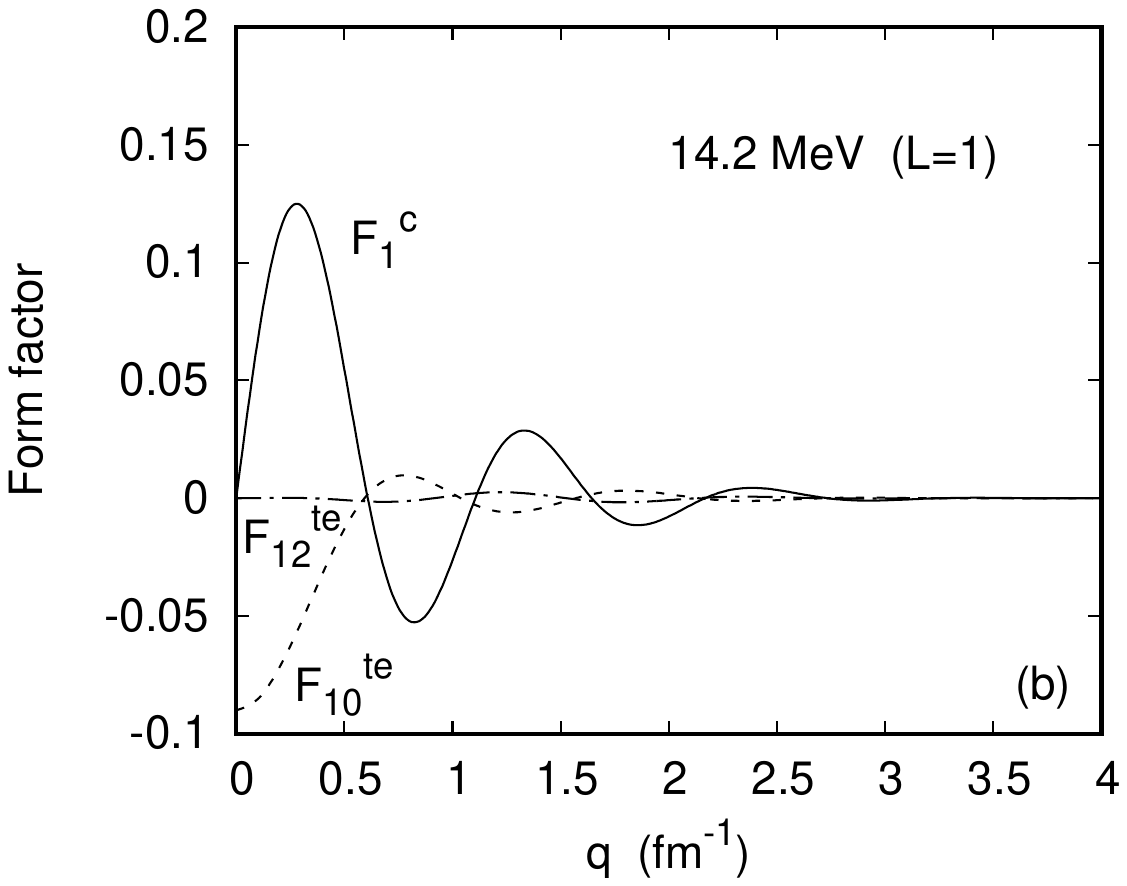}
\end{tabular}
\vspace{-1cm} 
\caption{
Form factors for the $1^-$ states in $^{208}$Pb at (a) 12.3 MeV and (b) 14.2 MeV as a function of momentum transfer $|\bfq|$. ---------, charge form factor $F_L^c;\;\,-\cdot -\cdot -,$ transverse electric form factor $F_{L,L+1}^{\rm te};\;\,-----$, transverse electric form factor $F_{L,L-1}^{\rm te}$. 
}
\label{fig2}
\end{figure*}

%\vspace*{-3.5cm} 

%Fig.3
\begin{figure*}[t]
\centering
\begin{tabular}{cc}
\hspace{-1cm}\includegraphics[width=.7\textwidth]{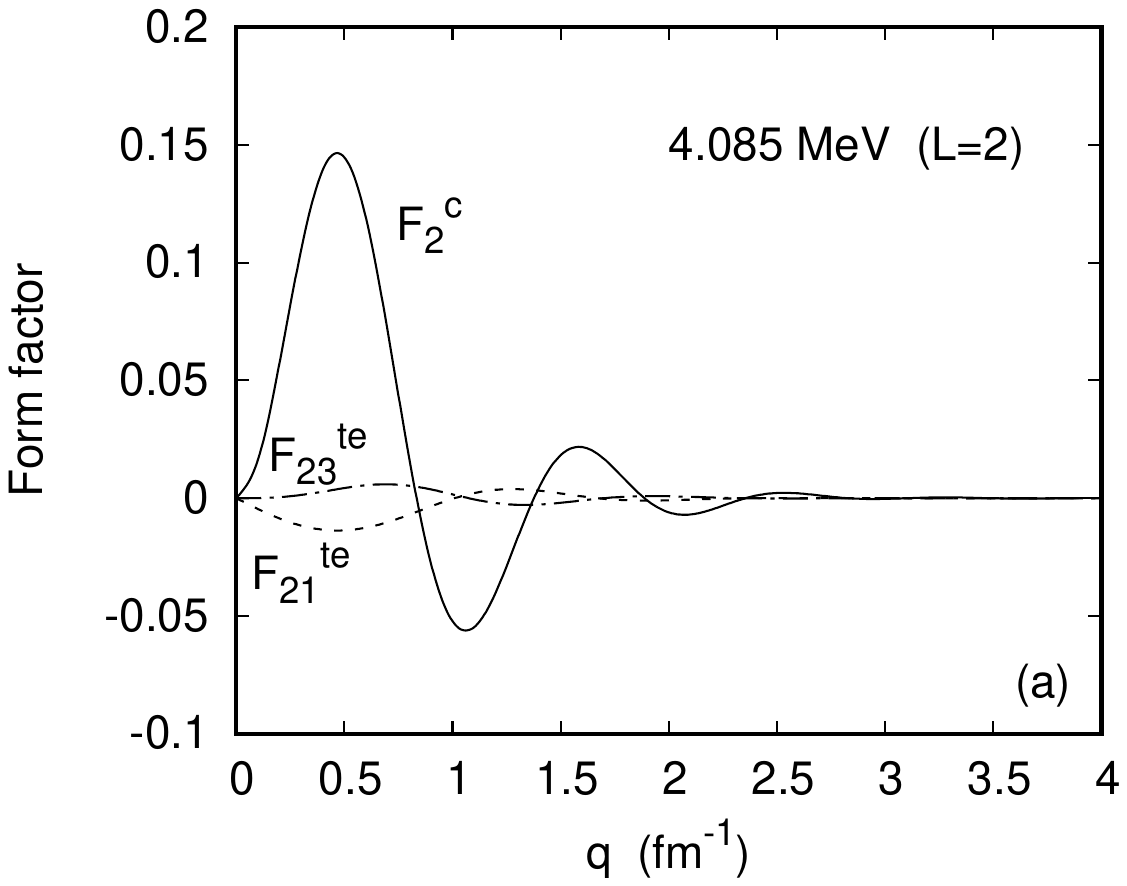}&
\hspace{-3.0cm} \includegraphics[width=.7\textwidth]{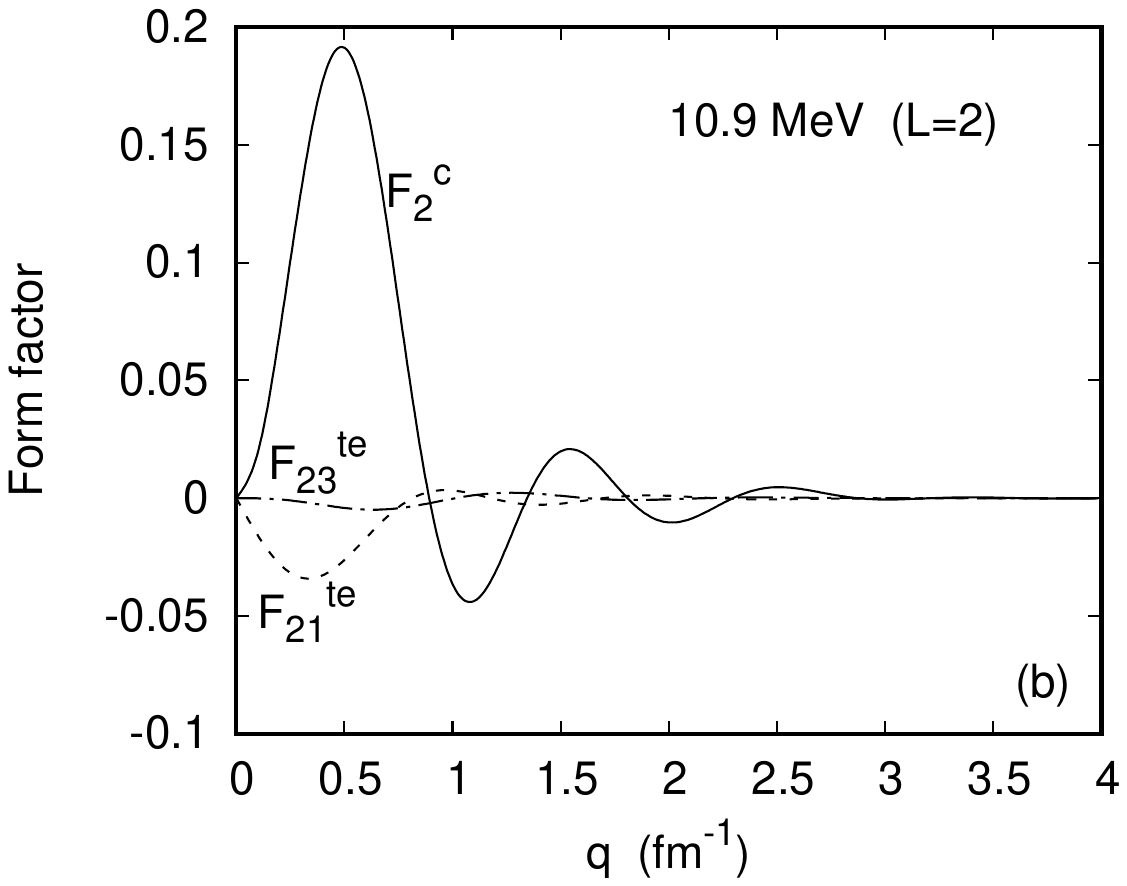}
\end{tabular}
\vspace{-1cm} 
\caption{
Form factors for the $2^+$ state in $^{208}$Pb at (a) 4.085 MeV and (b) 10.9 MeV as a function of momentum transfer $|\bfq|$. Legend as in Fig.2.
}
\label{fig3}
\end{figure*}

%\vspace{1.5cm} 

Introducing the transition charge density $\varrho_{L,\omega_L}$ and the current densities $J_{L\lambda,\omega_L}$ (with $\lambda=L\pm 1$ for the electric transitions considered here \cite{HB83})
as calculated from the SkP  HF+RPA  and changing to Fourier space in terms of the charge form factor $F_L^c$ and the transverse electric form factors $F_{L\lambda}^{te}$ defined by
$$F_{L,\omega_L}^c(q)\,=\,\int_0^\infty r^2\,dr\;\varrho_{L,\omega_L}(r)\;j_L(qr)$$
\begin{equation}\label{2.7}
F_{L\lambda,\omega_L}^{te}(q)\,=\,\int_0^\infty r^2\,dr\;J_{L\lambda,\omega_L}(r)\;j_\lambda(qr),
\end{equation}
with $j_L$ denoting spherical Bessel functions,
the multipole decomposition of the respective transition matrix elements can be written as \cite{GO62,RW87,Jaku22}
$$\langle LM,\omega_L|\,\hat{\varrho}(\bfq)\,|0\rangle\;=\;4\pi \,i^L \,F_{L,\omega_L}^c(|\bfq|)\;Y_{LM}^\ast(\hat{\bfq}),$$
\begin{equation}\label{2.8}
\langle LM,\omega_L|\,\hat{\bfJ}(\bfq)\,|0\rangle=-4\pi \,i \!\!\sum_{\lambda=L\pm 1}\!\! i^\lambda\,F_{L\lambda,\omega_L}^{te}(|\bfq|)\;\bfY_{L\lambda}^{\ast M}(\hat{\bfq}),
\end{equation}
for a target nucleus with spin zero in its ground state.
The angular functions $Y_{LM}$ and $\bfY_{L\lambda}^M$ are, respectively, spherical harmonic functions and vector spherical harmonics \cite{Ed}.
The form factors for a selection of excited states are displayed in Fig.~\ref{fig2} for $L=1$  and in Fig.~\ref{fig3} for $L=2$.

\vspace{0.2cm}

%Fig.4\\
\begin{figure}
\includegraphics[width=12cm]{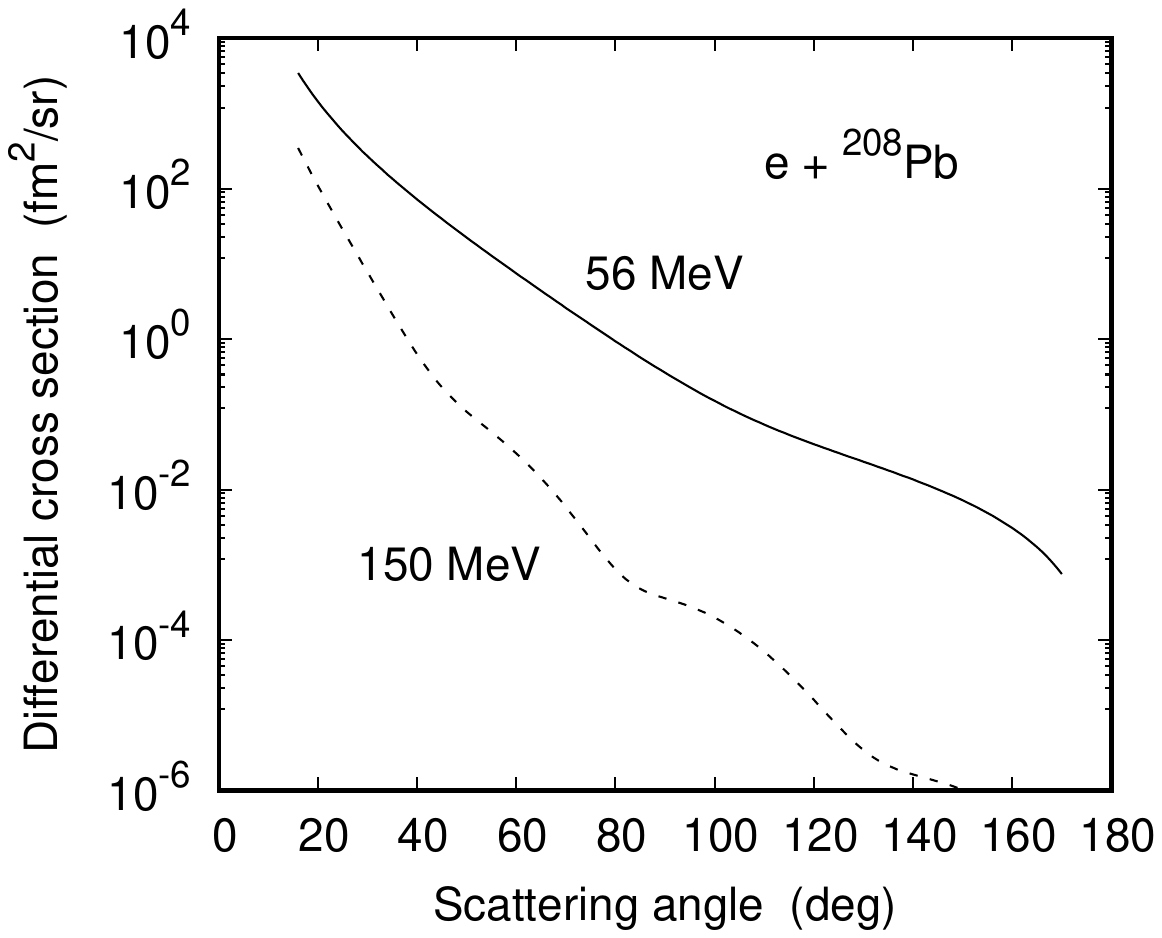}
\caption
{
Differential cross section $\frac{d\sigma_{\rm coul}}{d\Omega_f}$ for electrons of 56 MeV (---------)
and 150 MeV $(----)$ colliding with $^{208}$Pb as a function of scattering angle $\vartheta_f$.
}
\label{fig4}
\end{figure}

In order to split $A_{fi}^{\rm box}$ into a Coulombic part and a transverse electric (also termed magnetic) part, the gauge invariance is profited of by means of the decomposition \cite{Go61}
\begin{equation}\label{2.9}
\frac{g^{\mu \nu}}{q_i^2+i\epsilon}\;=\;-\;\frac{1}{\bfq_i^2}\,\delta_{\mu 0} \delta_{\nu 0}\;-\;\frac{\delta_{mn}-\hat{q_i}^m \hat{q_i}^n}{q_i^2+i\epsilon}\,\delta_{\mu m} \delta_{\nu n},
\end{equation}
where $g^{\mu\nu}$ is the metric tensor, $i=1,2$  and $\hat{q_i}^m =q_i^m/|\bfq_i|$ (with $m=1,2,3$) are 
the reduced components of $\bfq_i$.
Making use of
\begin{equation}\label{2.10}
T^{\mu\nu}(LM,\omega_L)\,=\,\sum_{\varrho,\tau} g^{\mu \varrho} g^{\nu\tau}T_{\varrho \tau}(LM,\omega_L)
\end{equation}
and inserting it together with (\ref{2.9}) into the formula (\ref{2.4}), it is seen that $A_{fi}^{\rm box}$ consists of four contributions,
$$A_{fi}^{\rm box}\,=\,\sum_{L,\omega_L}\left[ A_{fi}^c(L,\omega_L) + A_{fi}^{te1}(L,\omega_L)\right.$$
\begin{equation}\label{2.11}
+ \left. A_{fi}^{te2}(L,\omega_L)+A_{fi}^{te3}(L,\omega_L)\right].
\end{equation}
The Coulombic contribution $A_{fi}^c$ results from $\mu=\nu=0$ in (\ref{2.9}) for $q_1$ and $q_2$
and hence contains the denominator $\bfq_2^2\cdot \bfq_1^2.$
The transverse electric terms $A_{fi}^{te1}$ and $A_{fi}^{te2}$ originate from $\mu=\nu=0$ for $q_2$, respectively, for $q_1$. In the last term, $A_{fi}^{te3}$, one has $\mu,\nu \in \{1,2,3\}$ for $q_1$ and $q_2.$
Explicit expressions for these terms are provided in \cite{Jaku22}.  

\section{Results}

The leading-order contribution to elastic scattering is obtained from the phase-shift analysis for potential scattering \cite{Lan}. The Coulombic nuclear potential $V_T$ is derived from the $^{208}$Pb ground-state charge density distribution, given in terms of a 17-term Fourier-Bessel representation \cite{deV}.
The phase shifts are obtained by solving the corresponding radial Dirac equations with the help of the Fortran code RADIAL \cite{Sal}.
The determination of $f_{\rm coul}$ involves weighted summations of the phase shifts which are performed with the help of a threefold
convergence acceleration \cite{YRW}. This, however, leads to a divergent cross section at $0^\circ$ 
such that restriction is made to $\vartheta_f\geq 10^\circ$.
For the dispersion contribution 5 dipole states are taken into account, the lowest $1^-$ state at 5.512 MeV and 4 strong states in the giant dipole resonance region at 12.3, 14.2, 14.6 and 15 MeV.
In addition, 3 quadrupole states  are included, the lowest $2^+$ state at 4.085 MeV, the dominant isoscalar one at 10.9 MeV and one of the strong isovector ones at 21.6 MeV.
In addition the strong collective $3^-$ state at 2.615 MeV (which is the lowest excitation of the $^{208}$Pb nucleus) and an isovector state at 28.94 MeV are considered, omitting the weak higher $3^-$ isoscalar states.

The three-dimensional integrals required for the evaluation of $A_{fi}^{\rm box}$ are carried out numerically as described in \cite{Jaku22}.
The Coulombic term $A_{fi}^c$ is easily handled if the singularity of the electron propagator  (\ref{2.5}) in the polar integral is extracted analytically. 
Particular convergence problems are, however, encountered in the third magnetic term $A_{fi}^{te3}$,
where a logarithmic singularity arises when the pole from the electron propagator coincides with the singularity of one of the photon propagators (which occurs near $|\bfq|=2\omega_L/c$ for each state).
We note that a similar deficiency occurs  in the elastic second-order Born approximation \cite{FR74}. However, no singularity is present in the nonperturbative phase-shift analysis for potential scattering.  In fact it has been shown
that this logarithmic second-order Born singularity is cancelled by adding higher-order Born terms.
Therefore, the singularities in $A_{fi}^{te3}$ are considered as spurious, in particular because they affect only a small angular region (see, e.g. Fig.~\ref{fig6} in \cite{Jaku23}).
In the presentation of our results, $q$-values too close to the singularities are mostly avoided and a spline interpolation across the singularities is used instead. 

\subsection{Differential cross section}
\setcounter{equation}{0}

All results shown in this subsection include an average over the two initial-state spin projections,
\begin{equation}\label{3.1}
\frac{d\sigma}{d\Omega_f}\;=\;\frac12 \left( \frac{d\sigma}{d\Omega_f}(\bfzeta_i)\,+\,\frac{d\sigma}{d\Omega_f}(-\bfzeta_i)\right).
\end{equation}
Fig.~\ref{fig4} displays the phase-shift result for 56 MeV and 150 MeV electrons colliding with $^{208}$Pb.
While for 56 MeV the differential cross section decreases monotonously with angle, diffraction oscillations occur at the higher collision energy.

The dispersive correction to the Coulombic cross section is defined by 
\begin{equation}\label{3.2}
\Delta \sigma_{\rm box}\,=\, \frac{d\sigma_{\rm box}/d\Omega_f-d\sigma_{\rm coul}/d\Omega_f}{d\sigma_{\rm coul}/d\Omega_f}.
\end{equation}

\begin{figure}
\vspace{-1.0cm}
\includegraphics[width=12cm]{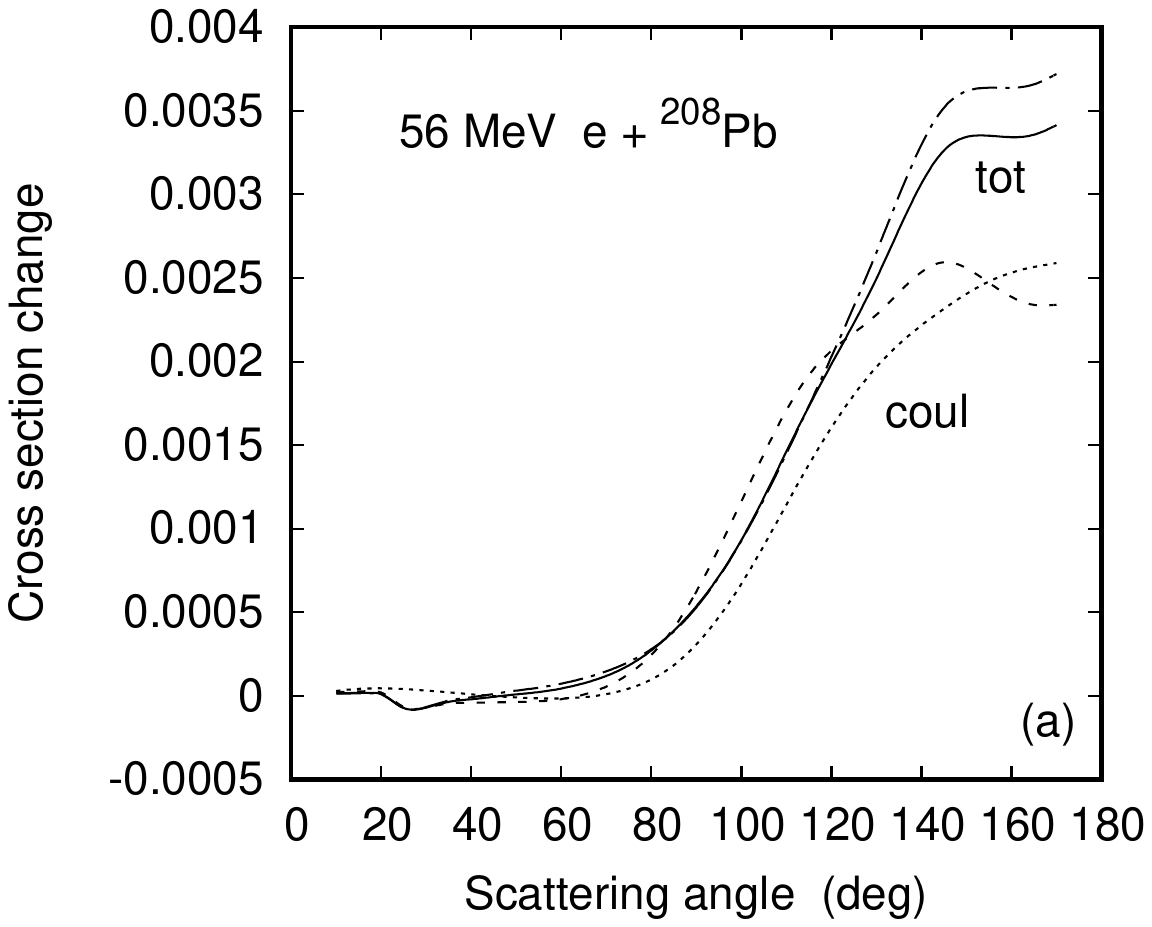}
%\vspace{-1.5cm}
\includegraphics[width=12cm]{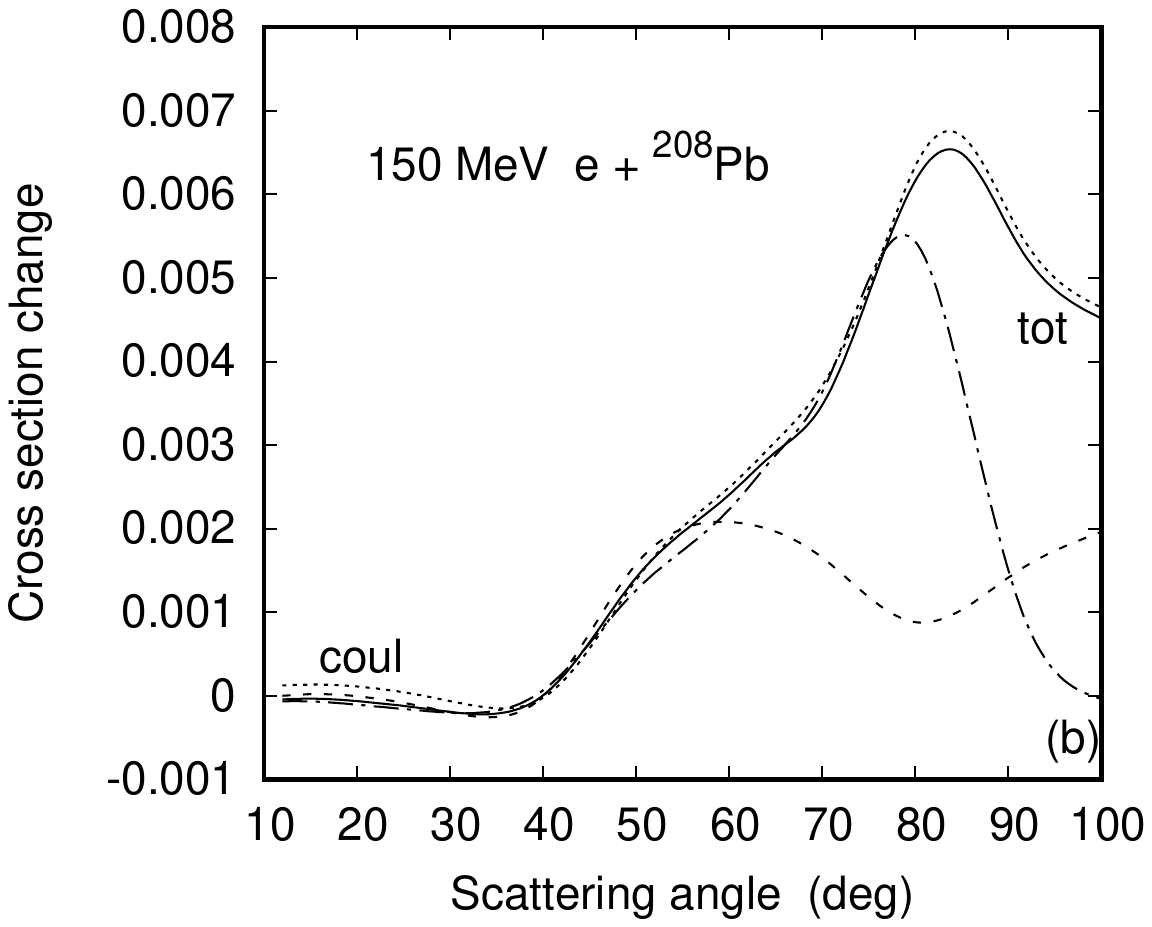}
\caption
{
Cross section change $\Delta \sigma_{\rm box}$ in (a) 56 MeV and (b) 150 MeV $e + ^{208}$Pb collisions as a function of scattering angle $\vartheta_f$.
Shown are the summed contributions from the $L=1$  states ($----$), from the $L\leq 2$ states $(-\cdot - \cdot -)$ and from the $L\leq 3$ states (---------).
Also shown is the Coulombic contribution pertaining to $A_{fi}^c$ for $L\leq 3 \;\,(\cdots\cdots$).
}
\label{fig5}
\end{figure}

From the linearity of the dispersive cross section (\ref{2.1}) in $A_{fi}^{\rm box}$, the contributions of the various excited nuclear states are additive, such that
\begin{equation}\label{3.3}
\Delta \sigma_{\rm box}\;=\;\sum_{L,\omega_L} \Delta \sigma_{\rm box}(L,\omega_L)
\end{equation}
is also additive in the cross-section change induced by a particular nuclear state characterized by $L$ and $\omega_L$.
The angular distribution of $\Delta \sigma_{\rm box}$ for collision energies of 56 MeV and 150 MeV is displayed in Fig.~\ref{fig5}.

%Fig.5\\

The total cross section change (\ref{3.3}) is very small, well below 0.5\% at 56 MeV, increasing to some extent  with energy. 
The dip at $30^\circ$ (Fig.~\ref{fig5}a), which is not present in the Coulombic contribution (pertaining to $A_{fi}^c$ in (\ref{2.11})), is an effect of the afore-mentioned logarithmic singularity for the states in the giant dipole resonance region.
Analyzing the influence of the  nuclear excitations with fixed $L$, it is seen that the main contribution in the forward hemisphere originates from the dipole states,
whereas at backward scattering angles the quadrupole states and the lowest octupole state become important.
There is a strong increase of $\Delta \sigma_{\rm box}$ with angle beyond $90^\circ$ at 56 MeV and beyond $35^\circ$ at 150 MeV.
This  onset corresponds to a momentum transfer of $|\bfq| \approx 0.4$ fm$^{-1}$ (if one approximates $|\bfq| \approx (2E_i/c)\sin(\vartheta_f/2)$, valid for $E_i \gg c^2$ and moderate recoil) where $F_1^c$ is peaked (Fig.~\ref{fig2}).

At the higher collision energy (Fig.~\ref{fig5}b), the total cross-section change is quite close to its Coulombic contribution in the whole angular range considered.
In contrast to the behaviour at 56 MeV, the contribution from the $L=3$ excitations becomes dominant beyond $80^\circ$.
This shows that for an increased momentum transfer the nuclear states of higher multipolarity come into play.

\subsection{Spin asymmetry}

The spin asymmetry is defined as the relative cross-section difference when the initial spin vector $\bfzeta_i$ is flipped.
For perpendicularly polarized electrons this leads to the Sherman  function $S$, which for Coulomb potential scattering without dispersion effects is given by
\begin{equation}\label{3.4}
S_{\rm coul}\,=\,\frac{d\sigma_{\rm coul}/d\Omega_f(\bfzeta_i) -d\sigma_{\rm coul}/d\Omega_f(-\bfzeta_i)}{d\sigma_{\rm coul} /d\Omega_f(\bfzeta_i)+d\sigma_{\rm coul}/d\Omega_f(-\bfzeta_i)}.
\end{equation}

%Fig.6\\
\begin{figure}
%\vspace{-1.5cm}
\includegraphics[width=12cm]{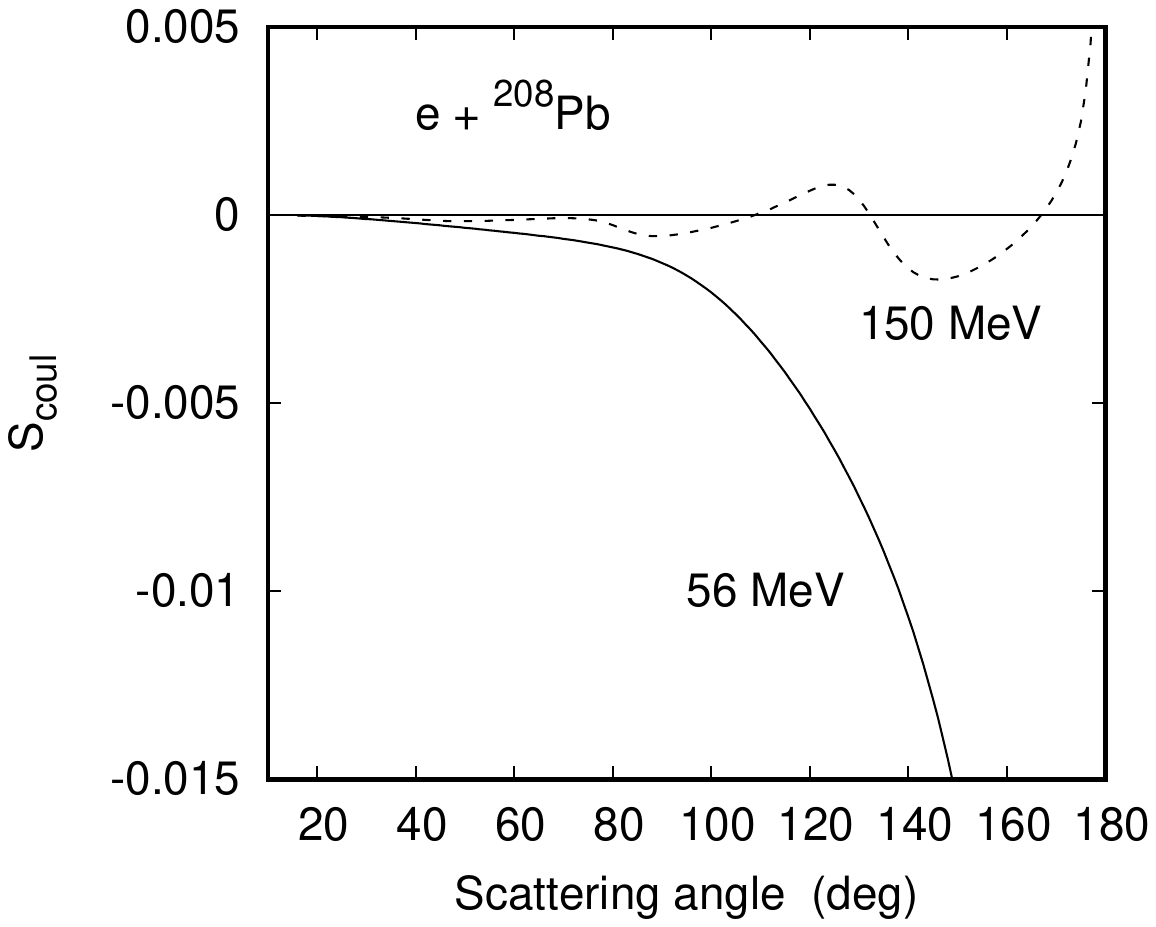}
%\vspace{-1.5cm}
\caption
{
Spin asymmetry $S_{\rm coul}$ for perpendicularly polarized electrons colliding with $^{208}$Pb at 56 MeV (--------) and 150 MeV ($----$) as a function of scattering angle $\vartheta_f$.
}
\label{fig6}
\end{figure}

It is shown in Fig.~\ref{fig6} for 56 MeV and 150 MeV.
The oscillations of $S_{\rm coul}$ at the higher collision energy, which are much more pronounced than those in the differential cross section, reveal that the spin asymmetry is more sensitive to details of the theoretical models than the spin-averaged cross section.
From the definition (\ref{3.4}) it follows that the extrema of $S_{\rm coul}$ correspond to the cross-section minima (see Fig.~\ref{fig4}).
At 150 MeV, $S_{\rm coul}$ has its maximum of $3.4\times 10^{-2}$ at $\vartheta_f=179.78^\circ.$
At 56 MeV where $d\sigma_{\rm coul}/d\Omega_f$ is structureless, $S_{\rm coul}$ decreases monotonously to its minimum of -0.47 at $179.4^\circ$.

When dispersion is included,  the spin asymmetry is calculated from 
\begin{equation}\label{3.5}
S_{\rm box}\,= \,\frac{d\sigma_{\rm box}/d\Omega_f(\bfzeta_i)-d\sigma_{\rm box}/d\Omega_f(-\bfzeta_i)}{d\sigma_{\rm box}/d\Omega_f(\bfzeta_i) + d\sigma_{\rm box}/d\Omega_f(-\bfzeta_i)}.
\end{equation}

In accordance with (\ref{3.2}), the relative change of the Coulombic Sherman function $S_{\rm coul}$ by dispersion is defined by
\begin{equation}\label{3.6}
d S_{\rm box}\,=\, \frac{S_{\rm box}-S_{\rm coul}}{S_{\rm coul}}.
\end{equation}
It should be noted that (\ref{3.6}) is not well defined in the vicinity of a zero of $S_{\rm coul}$, which for $^{208}$Pb occurs at an angle $\vartheta_f=16^\circ$ for 56 MeV and at $\vartheta_f=8.2^\circ,\;109^\circ,\;132^\circ$ and $167^\circ$ for 150 MeV collision energy (see Fig.~\ref{fig6}). 

\begin{figure*}[t]
\centering
\begin{tabular}{cc}
\hspace{-1cm}\includegraphics[width=.7\textwidth]{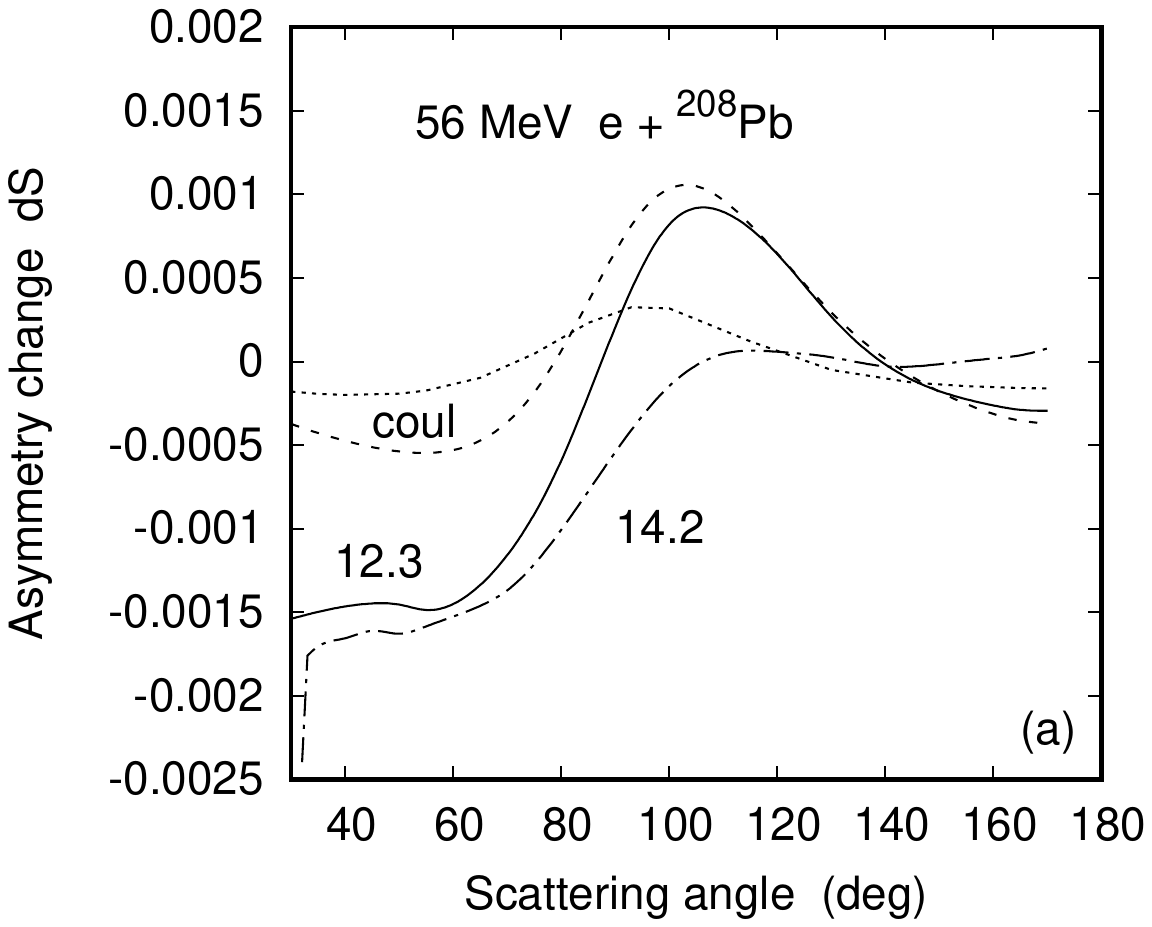}&
\hspace{-3.0cm} \includegraphics[width=.7\textwidth]{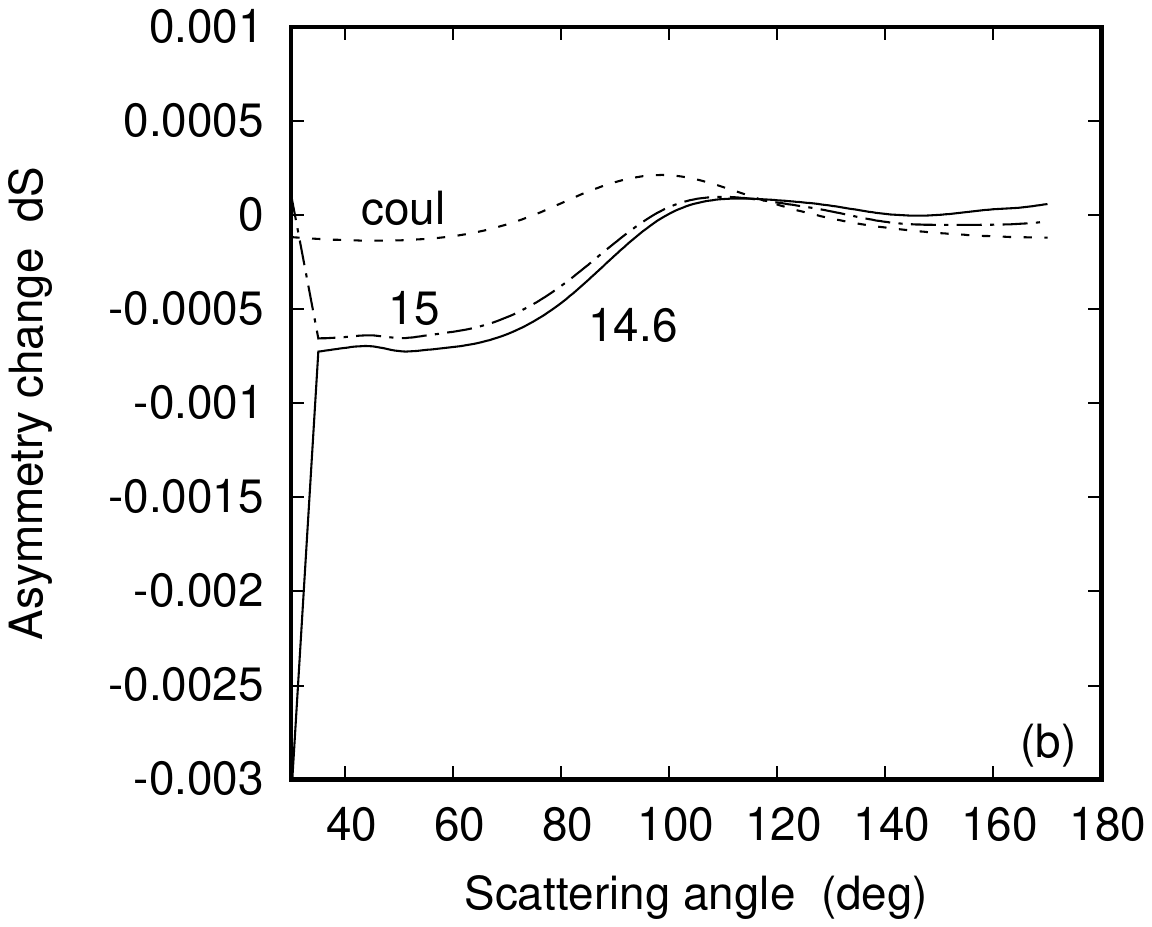}\\
\hspace{-1cm}\includegraphics[width=.7\textwidth]{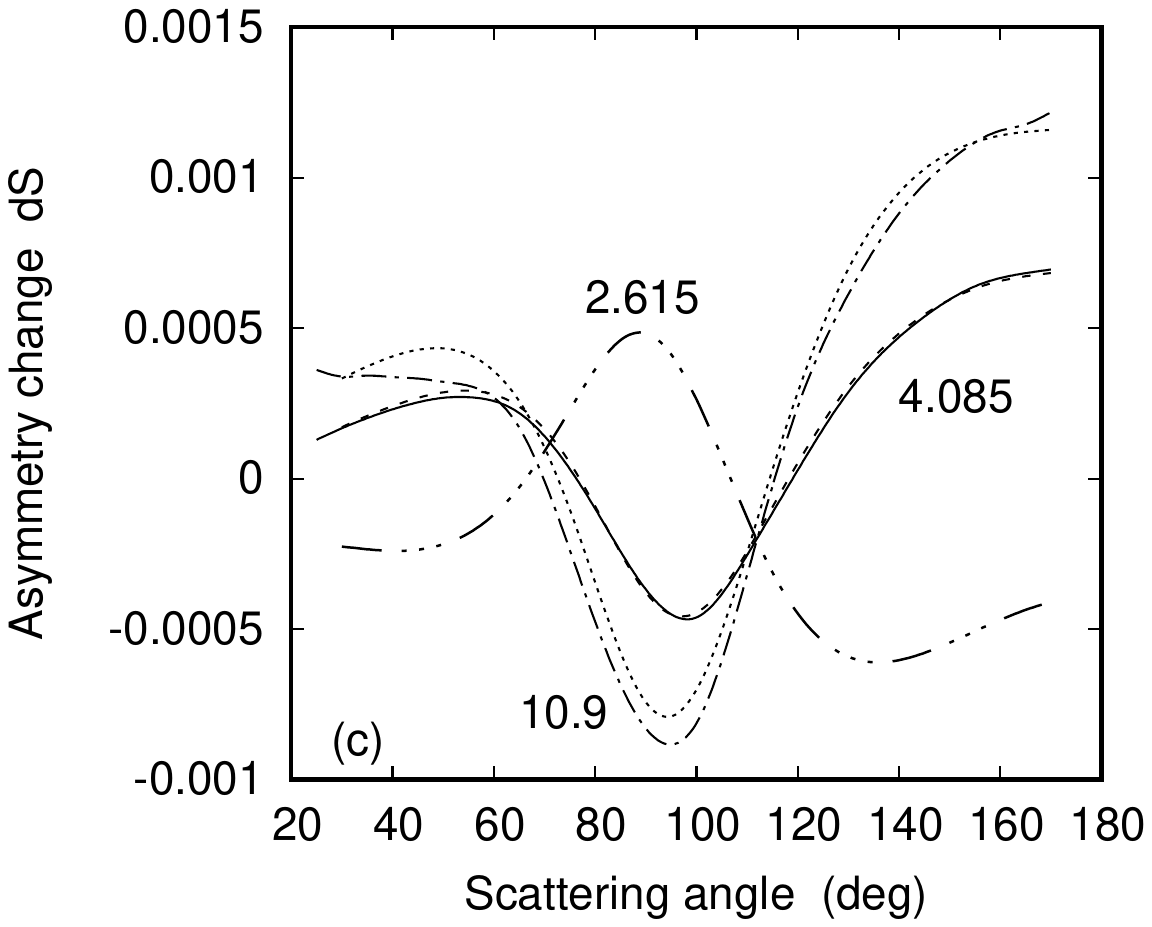}&
\hspace{-3.0cm} \includegraphics[width=.7\textwidth]{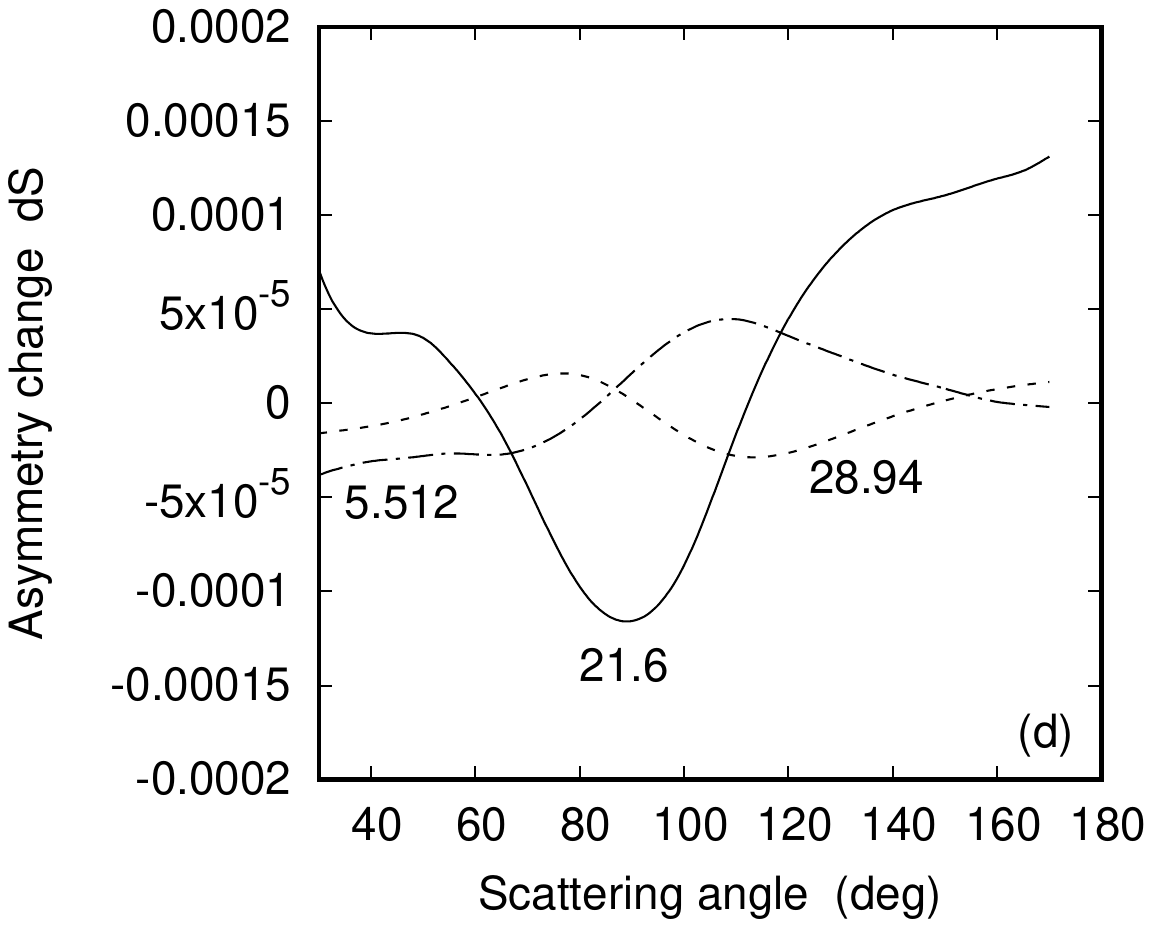}
\end{tabular}
\caption{
Change $dS_{\rm box}(L,\omega_L)$ of the Sherman function from 56 MeV $e + ^{208}$Pb collisions as a function of scattering angle $\vartheta_f$.
Shown in (a) are the results for the dipole states at 12.3 MeV (----------) and at 14.2 MeV $(-\cdot - \cdot -)$.
Included is the Coulombic contribution for the 12.3 MeV state ($----$) and for the 14.2 MeV state $(\cdots\cdots$).
Shown in (b) are the results for the dipole states at 14.6 MeV (--------) and at 15 MeV ($-\cdot -\cdot -)$. Included is the Coulombic contribution $(----)$ which coincides for the two states.
Shown in (c) are the results for the quadrupole states at 4.085 MeV (--------) and at 10.9 MeV $(-\cdot-\cdot-)$ as well as their Coulombic contribution ($----,\;4.085$ MeV; $\,\cdots\cdots$, 10.9 MeV).
Also displayed is the result for the octupole state at 2.615 MeV $(-\cdots -$) which coincides with its Coulombic contribution.
Displayed in (d) are the results for the $1^-$ state at 5.512 MeV $(-\cdot-\cdot -)$, for the $2^+$ state at 21.6 MeV (--------) and for the $3^-$ state at 28.94 MeV $(----)$.
}
\label{fig7}
\end{figure*}

For small dispersive cross-section changes as occur at low collision energies ($E_i \lesssim 150$ MeV), one has an approximate additivity of $S_{\rm box}$ and $dS_{\rm box}$ with respect to the contributions of the various nuclear intermediate states,
$$S_{\rm box}\,\approx\,\sum_{L,\omega_L} S_{\rm box}(L,\omega_L) -(N_L-1)\,S_{\rm coul},$$
\begin{equation}\label{3.7}
d S_{\rm box}\;\approx\;\sum_{L,\omega_L} d S_{\rm box}(L,\omega_L),
\end{equation}
where $N_L=10$ is the total number of the considered states.
The angular dependence of the contributions $dS_{\rm box}(L,\omega_L)$ is provided in Fig.~\ref{fig7}.
In order to elucidate the influence of magnetic scattering, the Coulombic contribution $dS_{\rm box}^{\rm coul}(L,\omega_L)$ (relating to $A_{fi}^c(L,\omega_L)$) is shown in addition.

Fig.~\ref{fig7}a displays the results for the two dominant dipole states at 12.3 MeV and 14.2 MeV excitation energy.
It is seen that for the 12.3 MeV state, one has basically Coulombic scattering in the backward hemisphere, but magnetic scattering at small angles which leads to a steep increase of $|dS_{\rm box}|$ with decreasing angle.
For the 14.2 MeV dipole state, magnetic scattering is important at all angles.
The behaviour of the two other high-lying dipole states at 14.6 and 15 MeV is quite similar (Fig.~\ref{fig7}b), their contribution to $dS_{\rm box}$ being mostly smaller. 

%Fig.7

The results for the two isoscalar $L=2$ states are shown in Fig.~\ref{fig7}c. It is seen that the 4.085 MeV state provides a Coulombic behaviour at nearly all angles, while the spin-asymmetry change by the 10.9 MeV state is considerably influenced by magnetic scattering at the angles below $60^\circ$.
Indeed, one may derive this behaviour from the form factors displayed in Fig.~\ref{fig3}b.
It is seen that the transverse electric form factor $F_{21}^{te}$ largely compensates the Coulombic $F_{2}^c$ up to $|\bfq| \approx 0.3$ fm$^{-1}$. 
At $E_i-c^2=56$ MeV, this corresponds to $\vartheta_f \lesssim 65^\circ$, in good accord with the diminished magnetic influence
on $dS_{\rm box}(L=2,\omega_L=10.9$ MeV) beyond $60^\circ$.
Included in Fig.~\ref{fig7}c is the  collective $L=3$ state at 2.165 MeV which is purely Coulombic and which contributes significantly to $dS_{\rm box}$ at the backward angles, together with the two $L=2$ states.

The low-lying dipole state at 5.512 MeV (Fig.~\ref{fig7}d), again strongly magnetic, as well as the isovector octupole states (like one of them, at 28.94 MeV) contribute negligibly to $dS_{\rm box}$.
The same is basically  true for the isovector quadrupole states (see the result for the 21.6 MeV state).

\begin{figure}
%\vspace{-1.5cm}
\includegraphics[width=12cm]{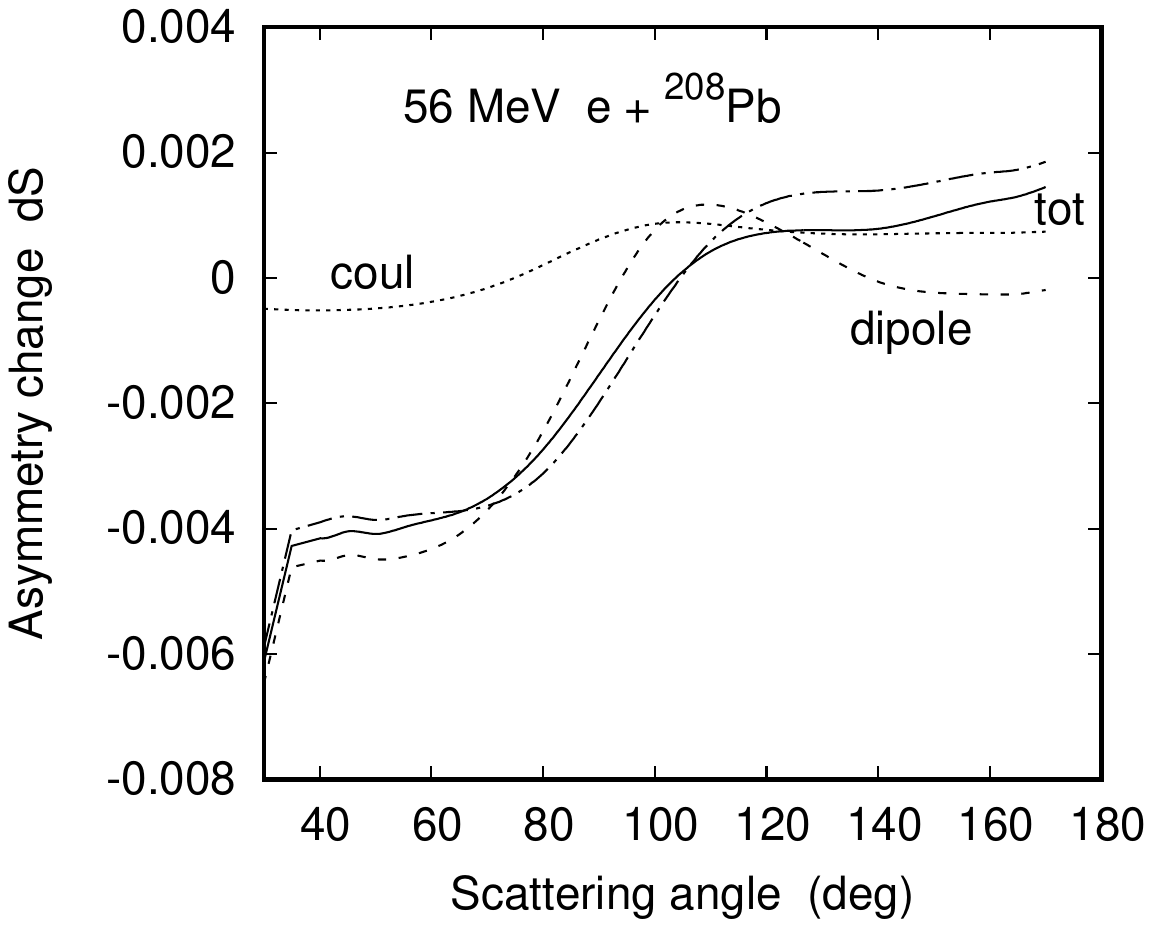}
%\vspace{-1.5cm}
\caption
{
Change $dS_{\rm box}$ of the Sherman function from  56 MeV $e+^{208}$Pb collisions as a function of scattering angle $\vartheta_f$.
Shown are the summed contributions from the dipole states
$(---)$, from the $L\leq 2$ states $(-\cdot - \cdot -$) and from the $L\leq 3$ states (--------).
Included is the Coulombic contribution for $L\leq 3$ $(\cdots\cdots)$.
}
\label{fig8}
\end{figure}

Fig.~\ref{fig8} displays the total spin-asymmetry change in comparison with the partial sums over $L=1$ and over $L\leq 2$ for the collision energy of 56 MeV.
Clearly, $dS_{\rm box}$ originates basically from the high-lying dipole states up to about $\vartheta_f=80^\circ$. 
Beyond that angle, the $L=1$ excitations become  less important and the isoscalar $L=2$ and eventually  $L=3$ states take over.
Included is the Coulombic contribution to $dS_{\rm box}$ which is close to zero in the whole forward hemisphere.

%Fig.8\\

\subsection{Comparison with the $^{12}$C target}

In the earlier work on $^{12}$C \cite{Jaku22,Jaku23} three  transient nuclear excitations were taken into account,   two strongly dominant dipole states in the giant resonance region at $\omega_L=23.5 $ MeV
and 17.7 MeV, and the lowest quadrupole excitation at 4.439 MeV.

%Fig.9\\
\begin{figure}
%\vspace{-1.5cm}
\includegraphics[width=11cm]{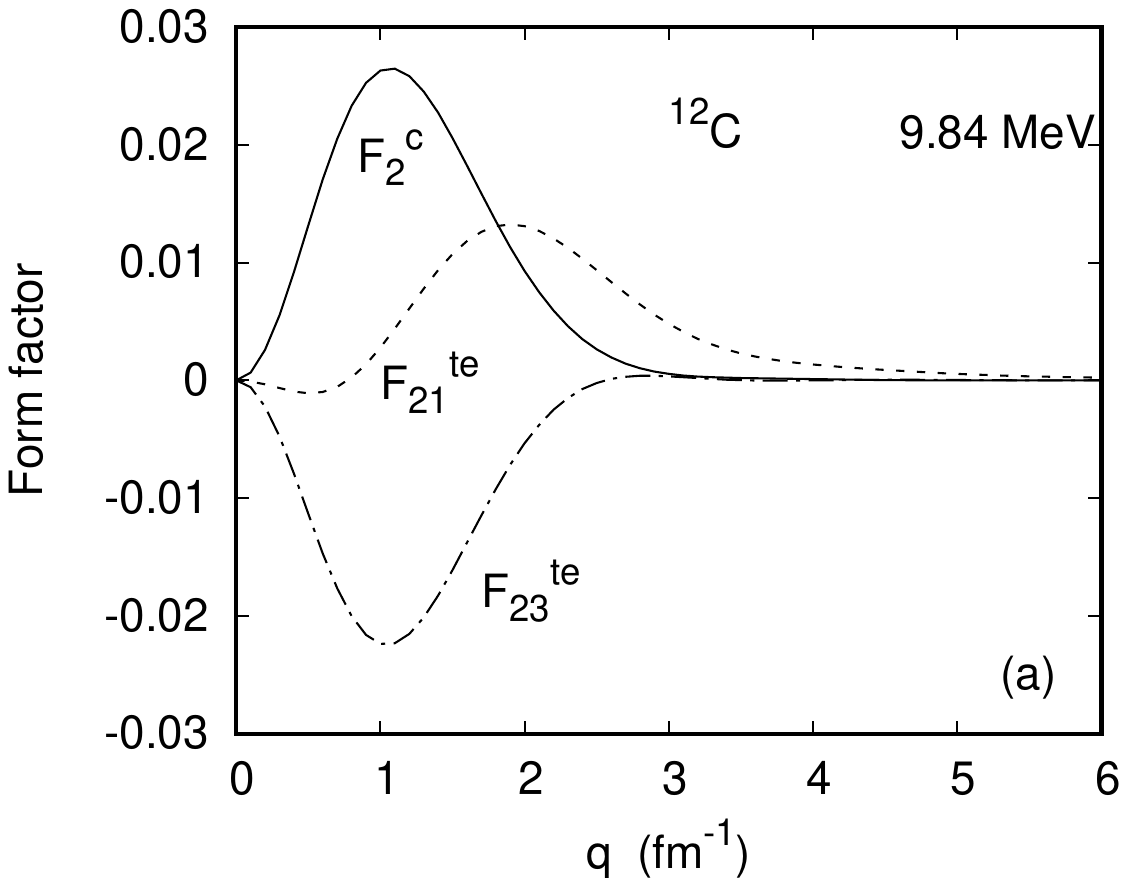}
%\vspace{-1.5cm}
\vspace{-0.5cm}
\includegraphics[width=11cm]{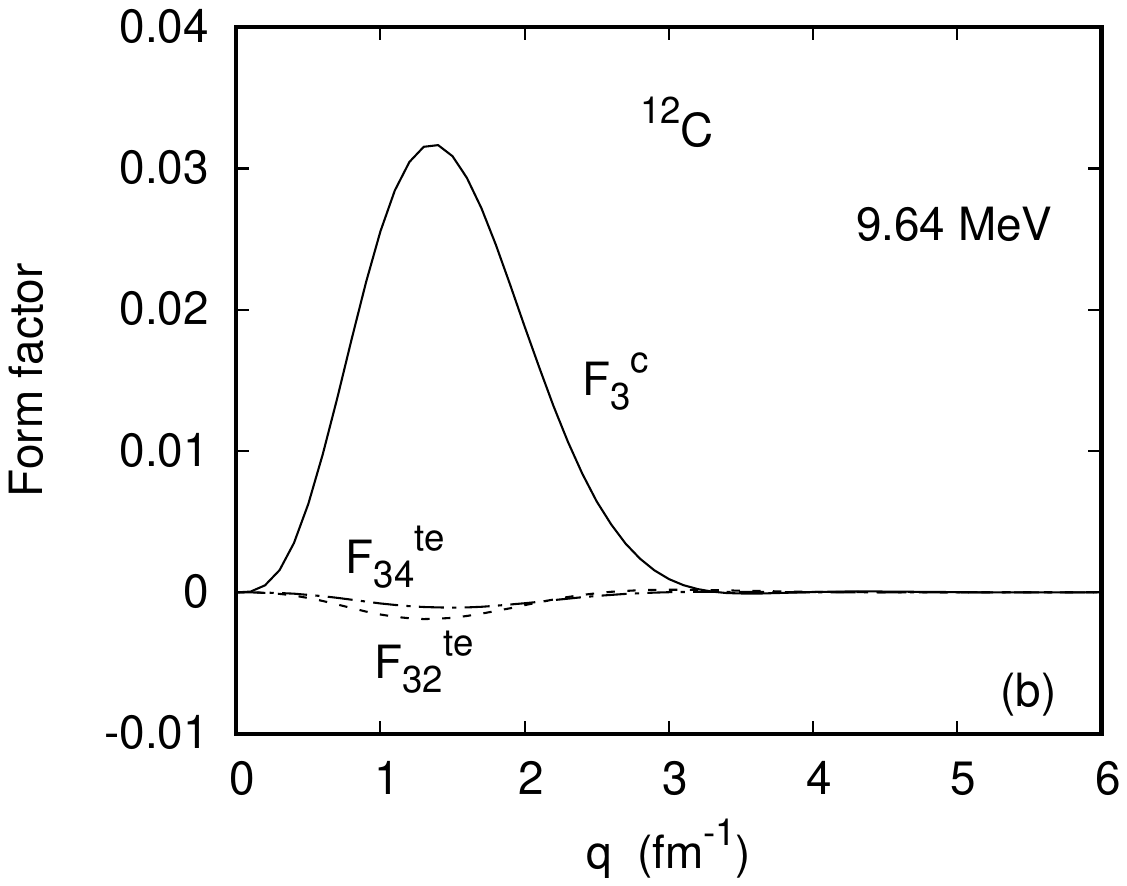}
\caption
{
Form factors for (a) the $2^+$ state at 9.84 MeV  and (b) the $3^-$ state at 9.64 MeV in $^{12}$C as a function of momentum transfer $|\bfq|.$ -----------, charge form factor $F_L^c$; $\,-\cdot -,$ transverse electric form factor $F_{L,L+1}^{\rm te}; \,----,$ transverse electric form factor $F_{L,L-1}^{\rm te}$.
}
\label{fig9}
\end{figure}

Guided by the present results for $^{208}$Pb, a few strong excitations with $L\leq 3$ are additionally taken into account. These are 
the second $2^+$ state at 9.84 MeV and also the lowest isoscalar $3^-$ state at 9.64 MeV as well as the strong isovector $3^-$ state at 14.8 MeV.  
The form factors for the  two lower states are plotted in Fig.~\ref{fig9}.
The form factors for the 14.8 MeV excitation are of a similar shape as those for the 9.84 state, but are smaller by a factor of 4.

\begin{figure}
%\vspace{-1.5cm}
\includegraphics[width=12cm]{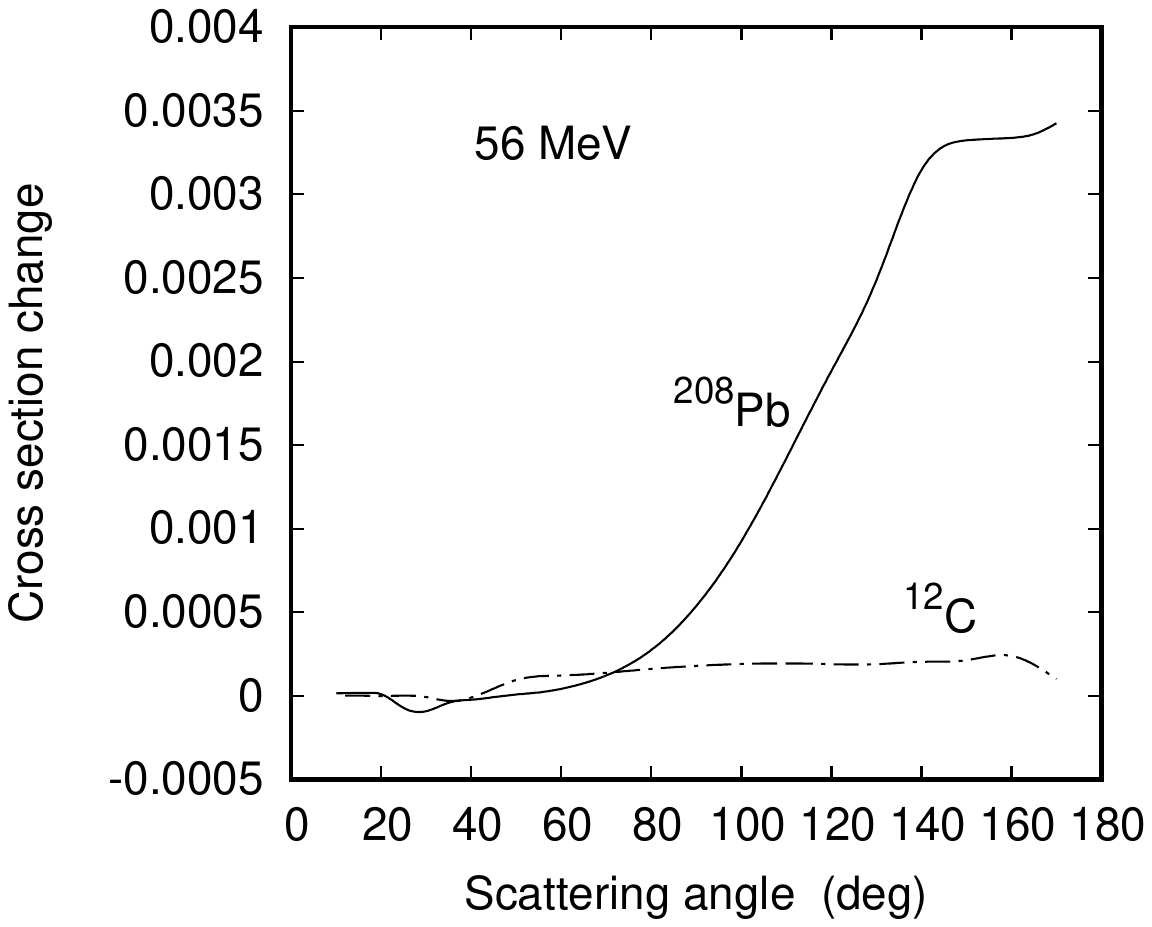}
%\vspace{-1.5cm}
\caption
{
Total cross section change $\Delta\sigma_{\rm box}$ for 56 MeV electrons colliding with $^{12}$C ($-\cdot -\cdot -)$ and with $^{208}$Pb (-------- ) as a function of scattering angle $\vartheta_f$. 
}
\label{fig10}
\end{figure}

\begin{figure}
%\vspace{-1.5cm}
\includegraphics[width=11cm]{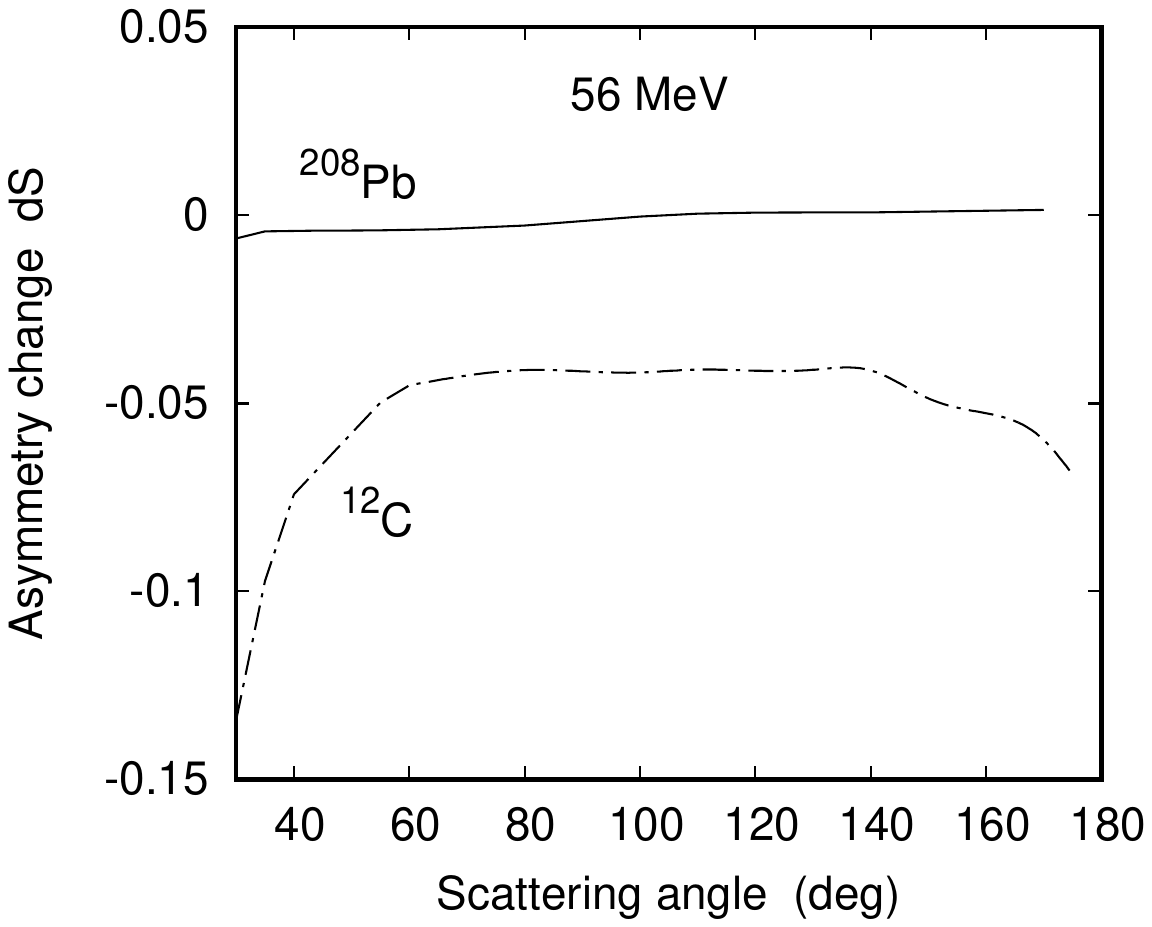}
%\vspace{-1.5cm}
\caption
{
Total change $dS_{\rm box}$ of the Sherman function for 56 MeV electrons colliding with $^{12}$C ($-\cdot - \cdot -$) and with $^{208}$Pb (-------) as a function of scattering angle $\vartheta_f$.
}
\label{fig11}
\end{figure}

It turns out, however, that the contribution of these additional states can safely be neglected. In fact, it is mostly well below 1\% for the spin-asymmetry change and a few percent for the cross-section change
at impact energies up to 70 MeV (increasing with energy to at most 10\% at 150 MeV).
The largest portion is provided by the $2^+$ state
because both its charge and transverse electric form factors are high (see Fig.~\ref{fig9}a), and we have included it in the results of Figs.~\ref{fig10}-\ref{fig11}.
The contribution of the octupole states is completely unimportant.

%Fig.10\\

In Fig.~\ref{fig10} the total dispersive cross-section change (summed over $L$ and $\omega_L$) for 56 MeV electrons colliding with a carbon and a lead target is shown.
While it is of similar magnitude for both targets below $70^\circ$,   the dispersion effects in the backward hemisphere are much larger for $^{208}$Pb than for $^{12}$C.

A corresponding behaviour is observed when at fixed angle (say, $50^\circ$) the impact energy is enlarged.
Due to the increase of momentum transfer, $\Delta \sigma_{\rm box}$ of $^{208}$Pb is eventually at 150 MeV a factor of 10 above the one of $^{12}$C.

Fig.~\ref{fig11} displays the results for the modification of the Sherman function by dispersion.
Its effect is at all angles larger for $^{12}$C than for $^{208}$Pb,
with a strong increase of $|d S_{\rm box}|$ at $\vartheta_f < 60^\circ$ which is not imitated for $^{208}$Pb.

%Fig.11\\

There are several reasons for the suppression of $d S_{\rm box}$ for a lead target (exemplified at 56 MeV and $\vartheta_f=40^\circ, \;160^\circ$).
Comparing the form factors of the dominant dipole excitation for  $^{12}$C and $^{208}$Pb, the shapes are similar, but the excitation energy is much smaller in $^{208}$Pb. Since high-lying dipole states are favoured, this might explain a reduction by a factor of up to 3.
However, more important is a basic sign change in $d S_{\rm box}$ when proceeding from $L$ to $L+1$.
Since for $^{208}$Pb the higher multipole states play a substantial role, there occurs a strong mutual cancellation between the contributions of different $L$.
In $^{12}$C, on the other hand, it is basically just the dipole states which are important at all momentum transfers considered.
A minor effect on the magnitude of $d S_{\rm box}$ results also from the phase difference between the Coulombic scattering amplitude $f_{\rm coul}^\ast$ and the one for dispersion in (\ref{2.1}).
This effect produces a slight reduction (up to 25\%) of the dispersive contribution to the cross section in the case of $^{208}$Pb as compared to $^{12}$C.

\section{Conclusion}

We have estimated the influence of dispersion on the elastic scattering cross section and on the beam-normal spin asymmetry from electron impact
on $^{208}$Pb by considering ten transiently excited states of multipolarity $L\leq 3$ within the second-order Born approximation.
We are confident that the inclusion of merely these states provides a reliable estimate of dispersion, since it turned out that the high-lying (isovector)
quadrupole and octupole states give a negligible contribution, and that the lowest $3^-$ state has a smaller effect than the most important $2^+$ state, except at very high momentum transfer.
As conjectured from distorted-wave Born as compared to (plane-wave) Born investigations for nuclear excitation, the use of the  second-order Born approximation for $A_{fi}^{\rm box}$ will lead to some underprediction of the dispersion effects. 
However, in case of the spin asymmetry it was shown that for momentum transfer $|\bfq| \lesssim 0.5$ fm$^{-1}$ (corresponding e.g. to  a collision energy of 56 MeV and $\vartheta_f \leq 180^\circ$)
Coulomb distortion effects are still small for a $^{208}$Pb nucleus \cite{Ko21}.

We obtained the result that, like for the $^{12}$C target, the excitations in the giant dipole resonance region supply the dominant part of dispersion in the forward hemisphere, both for the changes in the differential cross section and in the Sherman function.
This corroborates the model used in \cite{GH08,Ko21} for the foremost angles.
However, in contrast to $^{12}$C, dipole transitions lose their importance in the backward hemisphere.
In that region, quadrupole and at collision energies as high as 150 MeV even octupole transitions take over, with a dispersive cross-section modification mostly well below 1\% at all energies and angles considered.
The spin-asymmetry changes are larger, particularly at the smaller angles where they are in the percent region.

The importance of the excitation by the current-current interaction can be related to the size of the transverse electric form factors as compared to the charge form factor,
which varies strongly from state to state.
The total dispersive changes of the Sherman function are basically magnetic in the forward hemisphere,
and it is the transverse electric excitation which leads to the 
strong increase of the spin-asymmetry change when the scattering angle decreases. Even at backward angles, magnetic scattering is of some importance.
Concerning the cross-section changes, magnetic scattering
has a more moderate influence.

When comparison is made with the results for a $^{12}$C target, we have found that the cross-section changes are similar for the two targets at the smaller angles, while in the backward hemisphere they are one order of magnitude higher for lead than for carbon.
The reason, particularly at 150 MeV, is a constructive addition of the multipole contributions for $^{208}$Pb, such that $\Delta \sigma_{\rm box}$ is enhanced by the higer-multipole contibutions.

As concerns the change in the spin asymmetry, it is in fact much smaller for lead than for carbon at all angles.
This originates mainly from the destructive interference of the multipole contributions in $^{208}$Pb, since $dS_{\rm box}(L,\omega_L)$ basically switches sign when proceeding from $L$ to $L+1$. 
For $^{12}$C on the other hand, where the higher $(L\geq 2)$ multipoles are strongly suppressed, such cancellations do not occur.

In conclusion, we have verified the experimental observation that the influence of dispersion on the beam-normal spin asymmetry
is strongly suppressed for the $^{208}$Pb target,
even at the moderate collision energies considered here.
While the relevance of nuclear excitations decreases with impact energy \cite{Ad21},
a corresponding choice of smaller angles (as done in the experiments) may nevertheless reinstate their significance.
Hence a study of such type of mutual cancellations at larger energies may shed some more light into the current puzzle of the beam-normal spin asymmetry in this nucleus.

%\vspace{1cm}
%\pagebreak

%\noindent{\large\bf Acknowledgments}

%miau

\vspace{1cm}
%\pagebreak

\end{document}